\documentclass[aps,nofootinbib,onecolumn,showpacs,10pt]{revtex4-1}
\usepackage{graphicx,epsfig,amssymb}
\usepackage{epsf}
\usepackage{graphicx}

\newcommand\iden{\leavevmode\hbox{\small1\normalsize\kern-.33em1}}
\def \nn {\nonumber}
\def\w{{\rm w}}

\begin{document}
\vspace*{2cm}
\title{Pair production of single and double charged scalar pairs and their lepton flavor violating signals in the littlest Higgs model  at LHC }
\author{Ay\c se \c{C}a\u{g}{\i}l$^{1}$\footnote{ayse.cagil@cern.ch} and H\" useyin Da\u g$^{2}$\footnote{hdag@metu.edu.tr}}
\affiliation{\vspace*{0.1in} $^1$Physics Department, Middle East
Technical University, 06531 Ankara, Turkey\\
$^{2}$Middle East Technical University, North Cyprus Campus, Kalkanl\i, G\" uzelyurt, KKTC, Mersin 10, Turkey.}
\vspace*{1.0cm}
%
%
%
\begin{abstract}
In this work productions of charged and doubly charged scalars and their lepton flavor violating collider signals in the
framework of littlest Higgs model at LHC are studied. In the allowed
parameter region of the littlest Higgs model, the production rates of
the scalars of littlest Higgs model via $pp \to \phi^{++}\phi^{-}$,
$pp \to \phi^+\phi^-$ and $pp \to \phi^{++}\phi^{--}$ processes are calculated. We obtained that charged and doubly charged scalar pairs can be produced at LHC . Considering the possible
lepton flavor violating decays of charged scalars found in
literature, final state analysis is done. This analysis show that
depending on the model parameters, lepton number and lepton flavor
violations resulting from single and double charged heavy scalars of the littlest Higgs model can be observed at LHC.
\end{abstract}
\maketitle
\section{Introduction}
Up  to a certain energy level standard model(SM) has an impressive success explaining experimental data so far. 
Nowadays LHC is running with one of the main motivation of discovering SM Higgs boson and if SM Higgs boson to be discovered, 
the leading problem to focus will be stabilizing its mass against quadratically divergent radiative corrections, 
namely the hierarchy problem. Eventually the hierarchy problem is not the only problem that SM faces. 
SM also does not correctly account  for neutrino oscillations, thus nonzero masses of neutrinos, it can not explain dark matter phenomena, 
it has no explanation to strong CP problem, and many more. So it is obligatory to go beyond SM. 
So far with this purpose many theories and models have been introduced such as grand unified thories(GUT)\cite{GUT}, supersymmetric models(SUSY)\cite{SUSY}, 
models with extra dimensions\cite{ED}, $331$ models\cite{331mod}, left-right symmetric models\cite{LR}, (B-L) extended SM models\cite{BL}, doublet and triplet Higgs models\cite{DoubletM,TripletM}, Little Higgs(LH) models\cite{lh1,lh2,lhSimple,lhProduct} and many more. 
Each of these models have motivations to solve one or more of the problems that SM encounters.

Among these approaches Little Higgs models deserve attention due to their elegant solution to hierarchy problem. 
Little Higgs model was first proposed by Arkani-Hamed\cite{lh1}, and following the original idea several 
variations of LH models were introduced. Little Higgs models solve hierarchy problem by first enlarging the symmetry group of SM, 
and then by using collective symmetry mechanism to cancel out divergences of Higgs mass. The LH models differ in their assumed
 symmetry group and in the representation of scalar multiplets and they can be  classified into two groups as simple group and 
product group LH models\cite{lh1,lhSimple,lhProduct}. Among product group LH models, the most economical one is the Littlest Higgs Model\cite{lh1} 
which has a global $SU(5)$ symmetry containing  weakly gauged  $(SU2\otimes U1)^2$ subgroup. Littlest Higgs model as a consequence of its 
enlarged symmetry contains new heavy gauge bosons and a new heavy scalar sector arising from a complex $SU(2)$ group 
containing two neutral scalars, a charged scalar and a double charged scalar. The importance of this scalar sector is 
that especially charged scalars have very distinct collider signatures\cite{ays1,ays2,ays3,akeyord}.

The main problem the original Littlest Higgs model faced was satisfying electroweak precision data(EWPD) and to be 
consistent with the recent bounds on the lightest heavy scalar mass arising from searchs at Tevatron\cite{B2csaki,csaki1,perelstein2ew,B1rizzo,Bdawson,Bkilian,Bdias,Tevatron}. The 
free parameters of the model were strictly constrained meaning a severe fine tunning has to be done. In order to overcome 
this problem T parity were introduced which like R parity of SUSY introduces a discrete $Z2$ symmetry to the model\cite{Tparity}.
 As a consequence of implementing T parity interactions between SM particles and new particles are restricted, thus parameters 
were relaxed. Another consequence of this restriction is that the lightest new heavy gauge boson become a perfect candidate 
for dark matter\cite{Tparitydarkmatter}. But in order to account for non zero neutrino masses T parity is broken. Introducing 
T parity is not the only way of saving littlest Higgs model from strict constraints. Another method, which is also used in this work, 
is to charge fermions under both $SU(2)$ gauge groups\cite{B2csaki}.

As mentioned, one of the problems of SM is that it can not account for nonzero masses of neutrinos. The existence of complex $SU(2)$ scalar group
 in the littlest Higgs model allows neutrinos to gain their masses by implementing Majorana like mass term in the Yukawa Lagrangian without need of right handed neutrinos. The interactions of lepton doublets and complex $SU(2)$ scalars in Yukawa Lagrangian predicts lepton flavor and number violation by unit two, directly from decays of charged scalars, 
and this is an interesting and distinguishing feature of littlest Higgs model\cite{thanlept1,gaurlept1,cinlept_L2yue,gaurlept2}. Search for lepton flavor violating signals is one of the most interesting 
topics in collider physics, and in these searches one channel under investigation is the lepton flavor violation 
resulting from decays of charged and double charged scalars. Thus the models containing scalar , with hypercharge 
two are expected to give the most promising results. In litarature there are several models containing a scalar triplet 
and  lepton flavor violating signals in most of these models have been examined\cite{331}. Thus it will also be 
interesting to know the possible productions of charged scalars in littlest Higgs model and their lepton flavor violating 
signals at LHC.

In this work, we study the main production channels of heavy charged scalars and their lepton flavor violating collider signatures at LHC. In literature 
there are several works estimating sizable production rates of heavy charged scalars which are either model independent or arising from other models rather than littlest Higgs via 
$pp \to \phi^{++}\phi^{-}$,
$pp \to \phi^+\phi^-$ and $pp \to \phi^{++}\phi^{--}$ at hadron colliders\cite{EXmodelindependent}. In these searches they are basically investigating the possibility of a 
charged scalar satisfying the experimental bounds such as $M_\phi\geq150GeV$\cite{mphi}. In the littlest Higgs model due to the restrictions on the symmetry breaking scale, the heavy scalars cannot have a mass lower than $0.5TeV$. Thus these production channels including their lepton flavor and number violating final states need more investigation in the context of the littlest Higgs model. 
In proton collisions the charged scalars of the 
littlest Higgs model can be produced exculisively or in pairs. The exculisive production of the single charged scalar with a gauge boson mainly come from exchange of gauge bosons or neutral heavy scalars in s-channel. 
In the model couplings of the boson exchanges are dependent on $v'$(vacuum expectation value of the scalar triplet), thus when $v'$ is small enough to allow lepton flavor violation they are not observable. For the production of double charged scalars 
the exculisive channels $\bar q q' \rightarrow W^-_{L(H)}  \rightarrow W^+_L \phi^{--}$, $q q \rightarrow W^-_{L(H)} W^-_{L(H)} q'q' \rightarrow \phi^{--} q'q'$, 
$\bar q q' \rightarrow W^-_{L(H)}  \rightarrow W^+_H \phi^{--}$ and the pair production process $\bar q q' \rightarrow W^-_{L(H)}  \rightarrow \phi^+ \phi^{--}$ are analyzed in Ref. \cite{ref1Charged}. As also mentioned in Ref. \cite{ref1Charged}, the first three production processes are supressed by a factor of $v^{\prime 2}$, hence only the process $\bar q q' \rightarrow W^-_{L(H)}  \rightarrow \phi^+ \phi^{--}$ can have 
significant production rates at LHC. In addition to
these production processes, heavy charged scalars of the littlest Higgs model can be produced in pairs at LHC. 

In this work, the pair production processes: $pp \to \phi^{++}\phi^{-}$,
$pp \to \phi^+\phi^-$ and $pp \to \phi^{++}\phi^{--}$ via proton collisions at LHC are examined. In section two, we beriefly reviewed littlest Higgs model mostly following Ref. \cite{thanrev}, and we present the necessary formulation used 
in this work. In section three we present the production cross sections of production channels investigated in this work. We also present the final state analysis 
considering possible lepton flavor and number violating signals with our numerical estmates for LHC in section three. Finally at 
section four we concluded that 
through the all possible final states, three of them can be observed as lepton flavor or number violation at LHC. 
\section{Theoretical Framework}

The littlest Higgs model, as the most economical of the product group little Higgs models, have a global symmetry $SU(5)$ with a locally gauged
subgroup $\left(SU(2)\otimes U(1)\right)^2$. At a scale $\Lambda\sim 4 \pi f$, the global symmetry  $SU(5)$ is broken spontaneously to $SO(5)$ with a chosen vacuum
condensate. At low energies the dynamics of the symmetry breaking mechanism can be described by Nambu Goldstone boson (NGB) degrees of freedom, 
to each broken generator there exist a NGB. It is possible to represent the group structure as $1_0\oplus3_0\oplus 2_{1/2}\oplus 3_1$, where subscribes denote the hypercharge of the group. 
In the collective symmetry breaking mechanism, breaking of a global symmetry also triggers the symmetry breaking of the gauged subgroup of $\left(SU(2)\otimes U(1)\right)^2$ 
to $SU(2)\otimes U(1)$ of the standard model. During this symmetry breaking four of the NGBs are eaten by gauge bosons acquiring them their masses, while 
in the representationps a doublet $H$ and a triplet $\Phi$ remain physical. For these NGBs a Coleman Weinberg potential can be generated at one loop level. 
The generated scalar potential triggers usual electroweak symmetry breaking(EWSB) at chosen vacua of $v$ and $v'$ respectively for $H$ and $\Phi$. After 
EWSB, there are four new heavy scalars, $\phi^0$, $\phi^P$, $\phi^+$ and $\phi^{++}$ which remain in the scalar sector in addition to a light scalar $H$ which is identified as the SM Higgs boson. 
All scalars excluding $H$ are degenerate in mass:

\begin{eqnarray}
    M_\phi =\frac{\sqrt{2} f}{v\sqrt{1-(\frac{4 v'
    f}{v^2})^2}}M_H,
\end{eqnarray}
where $M_H$ is the mass of the Higgs boson, $f$ is the higher symmetry breaking scale of the littlest Higgs model, $v=246 GeV$ and $v'\leq \frac{v^2}{4f}$ are the vacuum expectation values (VEVs) of the Higgs field 
and the scalar triplet respectively which are bounded by electroweak precision data. 

The final particle content and properties of the gauge sector is dependent on the mixings of $U(1)$ and $SU(2)$ subgroups during the spontaneous breaking of $\left(SU(2)\otimes U(1)\right)^2$ to $SU(2)\otimes U(1)$. The mixing angles between the $SU(2)$
subgroups and between the $U(1)$ subgroups are defined respectively
as:
\begin{equation}\label{ssp}
    s=\frac{g_2}{\sqrt{g_{1}^2 + g_{2}^2 }}~~,~~~~~~~~ s^\prime=\frac{g'_2}{\sqrt{g_{1}^{\prime 2} + g_{2}^{\prime
    2}
    }}~~,
\end{equation}
where $g_{i}$ and
$g_{i}'$ are the corresponding couplings of $SU(2)_{i}$and
$U(1)_{i}$. After EWSB, gauge sector get additional mixings and mass terms resulting the final spectrum of gauge bosons. In the littlest Higgs model, gauge sector consists of heavy new gauge bosons 
$W_H^{\pm}$, $Z_H$, $A_H$ and light gauge bosons identified as SM gauge bosons; $W_L^{\pm}$, $Z_L$ and one massless boson $A_H$ identified as photon. 
The final masses of gauge bosons to the order of
$\frac{v^2}{f^2}$ are expressed as\cite{thanrev}:
\begin{eqnarray}\label{massesvectors}
M_{W_L^{\pm}}^2 &=&  m_w^2 \left[
    1 - \frac{v^2}{f^2} \left( \frac{1}{6}
    + \frac{1}{4} (c^2-s^2)^2
    \right) + 4 \frac{v^{\prime 2}}{v^2} \right], \nn \\
        M_{W_H^{\pm}}^2 &=& \frac{f^2g^2}{4s^2c^2}
    - \frac{1}{4} g^2v^2
    + \mathcal{O}(v^4/f^2)= m_w^2\left( \frac{f^2}{s^2c^2v^2}-1\right)
    ,\nn \\
    M_{A_L}^2 &=& 0 ,\nonumber \\
    M_{Z_L}^2 &=& m_z^2
    \left[ 1 - \frac{v^2}{f^2} \left( \frac{1}{6}
    + \frac{1}{4} (c^2-s^2)^2
    + \frac{5}{4} (c^{\prime 2}-s^{\prime 2})^2 \right)
    + 8 \frac{v^{\prime 2}}{v^2} \right],
    \nonumber \\
    M_{A_H}^2 &=&
    \frac{f^2 g^{\prime 2}}{20 s^{\prime 2} c^{\prime 2}}
    - \frac{1}{4} g^{\prime 2} v^2 + g^2 v^2 \frac{x_H}{4s^2c^2}
          = m_z^2 s_{\w}^2 \left(
    \frac{ f^2 }{5 s^{\prime 2} c^{\prime 2}v^2}
    - 1 + \frac{x_H c_{\w}^2}{4s^2c^2  s_{\w}^2} \right),
    \nonumber \\
    M_{Z_H}^2 &=& \frac{f^2g^2}{4s^2c^2}
    - \frac{1}{4} g^2 v^2
    - g^{\prime 2} v^2 \frac{x_H}{4s^{\prime 2}c^{\prime 2}}
    = m_w^2 \left( \frac{f^2}{s^2c^2 v^2}
    - 1 -  \frac{x_H s_{\w}^2}{s^{\prime 2}c^{\prime 2}c_{\w}^2}\right) ,
\end{eqnarray}
where  $m_w\equiv gv/2$, $m_z\equiv {gv}/(2c_{\w})$ and $x_H =
\frac{5}{2} g g^{\prime}
    \frac{scs^{\prime}c^{\prime} (c^2s^{\prime 2} + s^2c^{\prime 2})}
    {(5g^2 s^{\prime 2} c^{\prime 2} - g^{\prime 2} s^2 c^2)}$. In these
equations $s_{\w}$ and $c_{\w}$ are the usual weak mixing angles.

In littlest Higgs model, the free parameters are the symmetry braking scale $f$ and mixing angles $s$ and $s'$ and they are constrained by observables\cite{B2csaki,csaki1,perelstein2ew,B1rizzo,Bdawson,Bkilian,Bdias}. 
The data from Tevatron and LEPII constrain the mass of the lightest heavy scalar as $M_{A_H}\gtrsim900GeV$\cite{Bdawson,Tevatron}. In the original formulation of the littlest Higgs model, these data imposes strong constraints on symmetry breaking scale($f>3.5 - 4 TeV$). But in this work by gauging fermions in both $U(1)$ subgroups, fermion boson couplings are modified as done in\cite{B2csaki}. 
With this modification the symmetry breaking scale can be lowered to $f=0.75TeV(M_\phi\simeq0.5TeV)$ while mixing angles are restricted to be $s=0.8$ and $s'=0.6$, 
which allows the mass of the $A_H$ to be at the order of few $GeV$s with a large decay width. For larger values of $f$, the mixing agles are less restricted.

Finally, the fermions of the littlest Higgs model gain their masses through EWSB due to the Yukawa Lagrangian with an extended scalar sector. The additional scalar triplet of the model enables to 
implement a Majorano type mass term in Yukawa Lagrangian\cite{thanlept1,gaurlept1,cinlept_L2yue,gaurlept2}, such as:
\begin{equation}\label{lepviol1}
    {\cal L}_{LFV} = iY_{ij} L_i^T \ \phi \, C^{-1} L_j + {\rm h.c.},
\end{equation}
where $L_i$ are the lepton doublets $\left(
                                       \begin{array}{cc}
                                         l &\nu_l \\
                                       \end{array}
                                     \right)$,
and $Y_{ij}$ is the Yukawa coupling with $Y_{ii}=Y$ and $Y_{ij(i\neq
j)}=Y'$. Due to this term neutrinos gain mass without need of right handed neutrinos and also the lepton flavor violation arise from the decays of heavy scalars 
up to number of two. The values of Yukawa couplings $Y$ and $Y'$ are restricted
by the current constraints on the neutrino
masses\cite{neutrinomass}, given as; $M_{ij}=Y_{ij}v'\simeq
10^{-10}GeV$. Since the vacuum expectation value $v'$ has only an
upper bound; $v'<\frac{v^2}{4f}$, $Y_{ij}$ can be taken up to order of unity
without making $v'$ unnaturally small. In this work the values of
the Yukawa mixings are taken to be $10^{-10}\leq Y\leq 1$, 
$Y'\sim 10^{-10}$, and the vacuum expectation value $1GeV\geq v'\geq10^{-10}
GeV$.

While studying the production rates and final collider signals of littlest Higgs model scalars, their decay modes including the lepton flavor violating decays which are studied in T.Han et al\cite{thanlept1} are required. Due to their lepton flavor
violating modes, the total widths of the charged scalars will depend on the Yukawa
couplings $Y_{ii}=Y$ and $Y_{ij(i\neq j)}=Y'$. The decay modes and width of
$\phi^{++}$ are given as\cite{thanlept1}:
\begin{eqnarray}\label{dwp2}
 \nn   \Gamma_{\phi^{++}}&=&\Gamma (W^+_L W^+_L)+3 \Gamma ( \ell^+_i \ell^+_i) +3 \Gamma (\ell^+_i \ell^+_j )\\
    &\approx&  \frac{v^{\prime 2} M_{\phi}^3}{2 \pi v^4}+\frac{3}{8\pi } |Y|^2 M_\phi+\frac{3}{4\pi } |Y'|^2 M_\phi
\end{eqnarray}
For the single charged scalar, the decay modes and width are given
by\cite{thanlept1}:
\begin{eqnarray}\label{dwp1}
\nn \Gamma_{\phi^+}&=&3 \Gamma ( \ell^+_i \bar\nu_i)+6 \Gamma
(\ell^+_i \bar\nu_j)+\Gamma ( W_L^+ H)+\Gamma
(W_L^+ Z_L)+\Gamma (t \bar{b})+\Gamma (T \bar{b})\\
 &\approx & \frac{N_c M^2_t M_\phi }{32 \pi f^2}+\frac{v^{\prime
2}M^3_\phi }{2\pi v^4} +\frac{3}{8\pi
    }
    |Y|^2 M_\phi+\frac{3}{4\pi
    }
    |Y'|^2 M_\phi.
\end{eqnarray}
In this final expression, the decay of single charged scalar to $T\bar{b}$ is neglected since in the parameter space considered in this work, $M_\phi \sim M_T$, hence this decay is suppressed. It is seen from the decay widths of the scalars that lepton number
violation is proportional to  $|Y|^2$ if the final state leptons are
from the same family and to $|Y'|^2$ for final state leptons are
from different generations.


%
\section{The results and discussions}

The scattering amplitudes of the processes $pp \to \phi^{++}\phi^{-}$,
$pp \to \phi^+\phi^-$ and $pp \to \phi^{++}\phi^{--}$ depend on the parameters $s$, $s'$ and $f$, as well as center of mass(cms) energy $\sqrt{S}$ of LHC. While 
dependence on the mixing angles is weak, the dependence on the symmetry breaking scale $f$ and so on the mass of the new scalars is significant. In this work, we first examined the dependence
 of the production cross sections of the charged pairs to $M_\phi$ for different values of $\sqrt{S}$. In our calculations we have chosen $s=0.8$ and $s'=0.6$ allowed by the precision data. The symmetry breaking 
$f$ is taken in the range $0.75TeV$ to $3TeV$, thus the corresponding values for the mass of the heavy scalars vary in between $0.5TeV$ to $2TeV$. In our numerical calculations we have taken 
$M_H=120\pm3 GeV$, $M_{Z_L}=91.188\pm0.002 GeV$, $M_{W_L}=80.40\pm0.02GeV$ and $s_W=0.47$\cite{pdg}.

\begin{figure}[h]
\begin{center}
\includegraphics[width=9cm]{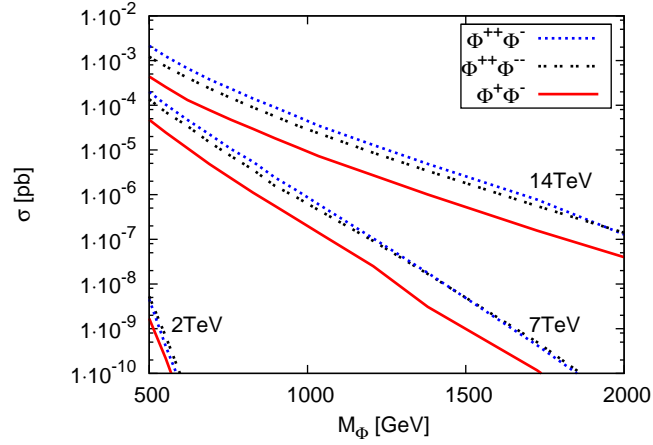}
\end{center}
\caption{The production cross sections of the scalar pairs with respect to $M_\phi$ for $\sqrt{S}=2TeV$, $\sqrt{S}=7TeV$ and $\sqrt{S}=14TeV$ when $s=0.8$ and $s'=0.6$.}
\label{fig::1}
\end{figure}

In figure \ref{fig::1}, we have plotted the production cross sections of the scalar pairs with respect to $M_\phi$ for $\sqrt{S}=2TeV$, $\sqrt{S}=7TeV$ and $\sqrt{S}=14TeV$ when $s=0.8$ and $s'=0.6$. It is seen from Fig.\ref{fig::1} that 
the productions of heavy scalar pairs are not observable at $\sqrt{S}=2TeV$, since their rates are at the order of $10^{-9}pb$. At $\sqrt{S}=7TeV$, the production rates of the scalar pairs are in the order of $10^{-4}pb$, which is in the reach for LHC. For the case 
$M_{\phi}=0.5TeV$ and $\sqrt{S}=14TeV$ scattering amplitudes for the processes $pp \to \phi^{++}\phi^{-}$,
$pp \to \phi^+\phi^-$ and $pp \to \phi^{++}\phi^{--}$ reach to values $2.9\times10^{-3}pb$, $0.5\times10^{-3}pb$ and $1.2\times10^{-3}pb$ respectively. Thus if LHC reaches to an integrated luminosity of $100fb^{-1}$, which is planned to be achieved within two years\cite{LHClum}, up to hundreds of $\phi^{++}\phi^{--}$ and 
$\phi^{++}\phi^{-}$ pairs can be produced. On the other hand number of $\phi^{+}\phi^{-}$ production can not exceed 50 events even in maximal conditions. Due to these 
predictions, for the final state analysis we concentrate on the production channels $pp \to \phi^{++}\phi^{-}$ and $pp \to \phi^{++}\phi^{--}$.

\begin{figure}[tbh]
\begin{center}
\includegraphics[width=7cm]{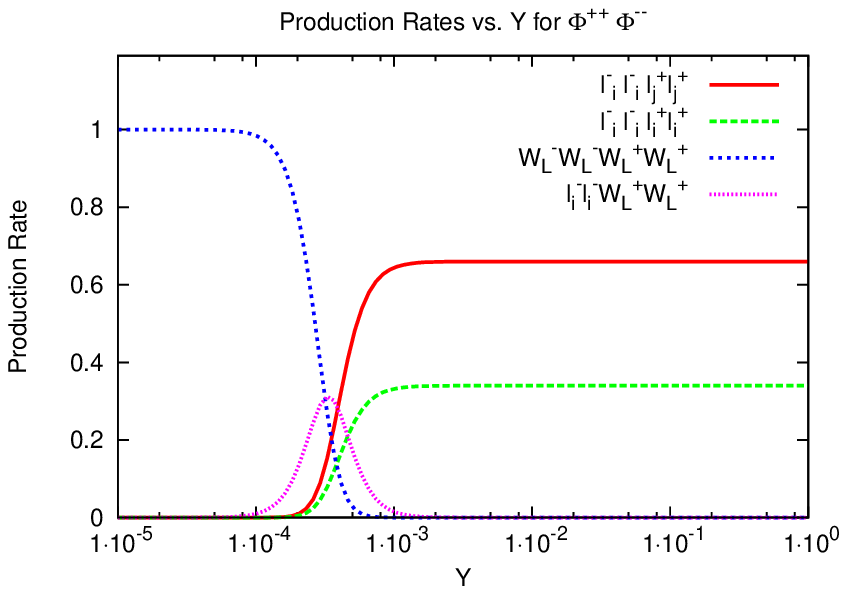}\hskip0.2cm\includegraphics[width=7cm]{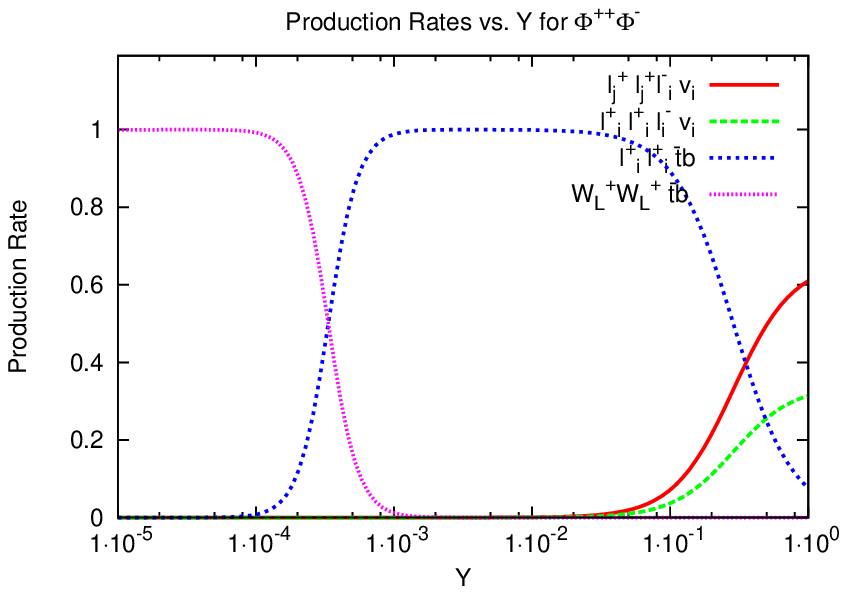}
\vskip2mm
\hskip1cm (a)\hskip5cm (b)
\end{center}
\caption{Production rates of final decay modes of $\phi^{++}\phi^{--}$(a) and $\phi^{++}\phi^{-}$(b) vs $Y$.}
\label{fig::decays}
\end{figure}

At LHC, heavy charged scalars of the littlest Higgs model will be identified from their lepton flavor violating decay modes\cite{akeyord,thanlept1,gaurlept1,thanrev,gaurlept2,aysbook}. As stated in Eqs. \ref{dwp2} and \ref{dwp1}, the decay modes of the charged scalars are strongly dependent on 
Yukawa couplings $Y$ and $Y'$, thus to the VEV of the scalar triplet, $v'$. By choosing $v'$ in the range $1GeV>v'>10^{-10}GeV$, the Yukawa couplings are chosen as $1>Y>10^{-10}$ and $Y'\simeq 10^{-10}$. For this range, we plotted the 
final decay modes of charged pairs $\phi^{++}\phi^{--}$ and $\phi^{++}\phi^{-}$ in figure \ref{fig::decays}.

For $\phi^{++}\phi^{--}$ pair, it is seen from Fig. \ref{fig::decays}(a) that for $Y<10^{-4}$ both of the double charged scalars decay into SM particles. The final state in this case will be 
two $W_L$ pairs. Due to low production rates and huge background at LHC, this case has no significance. For $Y>10^{-3}$ both scalars will
 decay into leptons violating lepton flavor. In this case $2/3$ of the final stay will be $l_il_i \bar{l}_j \bar{l}_j$ where $i$ and $j$ stand for different family of leptons, and the lepton flavor is explicitly violated by four giving a collider signal of four leptons consist of two same sign leptons from same family. This channel is free from any SM backgrounds. 
Also for $Y>10^{-3}$, $1/3$ of the double charged pairs wil decay into $l_il_i\bar{l}_i\bar{l}_i$. Even if these decays happen via lepton flavor violating decays of 
double charged scalars, the final signal will be indistinguishable from four leptons coming from SM backgrounds. For the range $10^{-3}>Y>10^{-4}$, one of the double charge scalar decay 
into SM bosons, and the other one decay into same sign leptons of same family, thus the final signal will be 
$W^-_L W^-_L \bar{l}_i \bar{l}_i$. In the view of the collider observables, this scenario is the most interesting because 
it will yield lepton number violation by two. Since $W_L$ decay into jets with a branching ratio of $0.6$, $36\%$ 
of the final states resulting from $W^-_L W^-_L \bar{l}_i \bar{l}_i$ will be two leptons of same sign 
and same family acompanied by jets, violating lepton number by two and free from any backgrounds at LHC.

For $\phi^{++}\phi^{-}$, we plotted production rates of the final signals depending on $Y$ in Fig. \ref{fig::decays}(b). For 
$Y<10^{-3}$, both scalars will decay into SM particles and the final mode will be $W_L^+W_L^+\bar{t}b$. Due 
to low production rates of charged scalars and huge SM background, this case is not promising at LHC. 
For $Y$ close to unity ($Y\simeq1$),both scalars decay into leptons. In this case $2/3$ of final states will be 
$\bar{l}_i\bar{l}_i l_j v_j$ and $1/3$ of the final states $\bar{l}_i\bar{l}_i l_i v_i$. In this scenario, 
$\bar{l}_i\bar{l}_i l_j v_j$ final states will be violating lepton flavor, which is a significant observable at LHC free from any SM backgrounds. For $\bar{l}_i\bar{l}_i l_i v_i$, since all of the leptons are from same family, the final states cannot be distinguished from SM backgrounds. 
For $\phi^{++}\phi^{-}$ the most interesting final state can be observed when $10^{-3}<Y<1$. In this case the double charged scalar will decay into two leptons of same sign and family and the single charged scalar will decay into jets. In this case the final collider signal will be $\bar{l}_i \bar{l}_i \bar{t}b$, 
two same sign and family leptons plus jets, violating lepton number by two and free from any backgrounds. 

After investigating the production rates of $\phi^{++}\phi^{--}$ and  $\phi^{++}\phi^{-}$ pairs and analyzing the final states of the processes, we combine the results to find possible lepton flavor and number violating final events at LHC. The final states are free from SM backgrounds and they are listed as:
\begin{itemize}
 \item $l_il_i\bar{l}_j\bar{l}_j$, four lepton final states: These final states are coming from the decays of $\phi^{++}\phi^{--}$. For $Y>10^{-3}$, $2/3$ of double charged pairs will decay in four leptons, with lepton flavor violation by four. In this final state  double charged scalars can be reconstructed from same sign and same family lepton pairs.
 \item $\bar{l}_i\bar{l}_i l_j v_j$, three leptons plus missing energy: For $Y\simeq1$, these final states are coming from the decays of $\phi^{++}\phi^{-}$. $2/3$ of the scalar pairs will decay into two same sign same family leptons plus one additional lepton with opposite sign from another family and the missing energy of the neutrino. In this final state the observed lepton flavor violation will be two since the family of the neutrino can not be identified.  
 \item $\bar{l}_i\bar{l}_i \bar{t} b$, two leptons plus jets: This is the most interesting final state arising from the decays of scalar pairs because the resulting signal violates lepton number by two. This final state can be observed from the final decays of $\phi^{++}\phi^{-}$ when $10^{-1}>Y>10^{-3}$, and also from the semileptonic decays of $\phi^{++}\phi^{--}$ if two of the final state $W_L$'s decay into jets when $10^{-3}>Y>10^{-4}$. 
\end{itemize}

 \begin{figure}[h]
\begin{center}
\includegraphics[width=9cm]{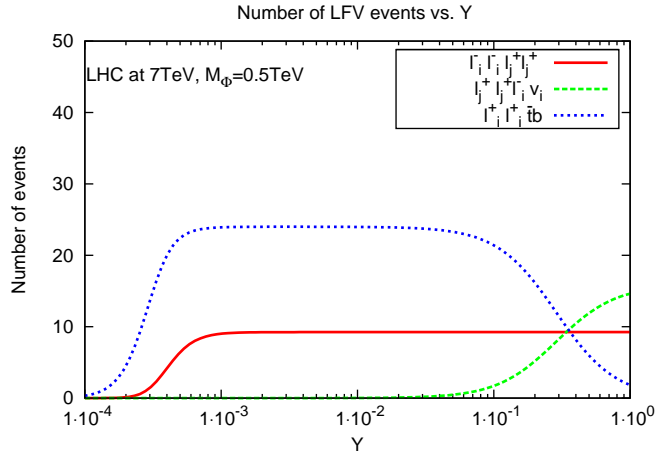}
\end{center}
\caption{Dependence of total number of lepton number and lepton flavor violating final states from decays of scalar pairs at LHC when $\sqrt{S}=7TeV$ and $M_\phi=0.5TeV$ for an integrated luminosity of $100fb^{-1}$ on Yukawa coupling $Y$.}
\label{fig::No7}
\end{figure}

 \begin{figure}[h]
\begin{center}
\includegraphics[width=9cm]{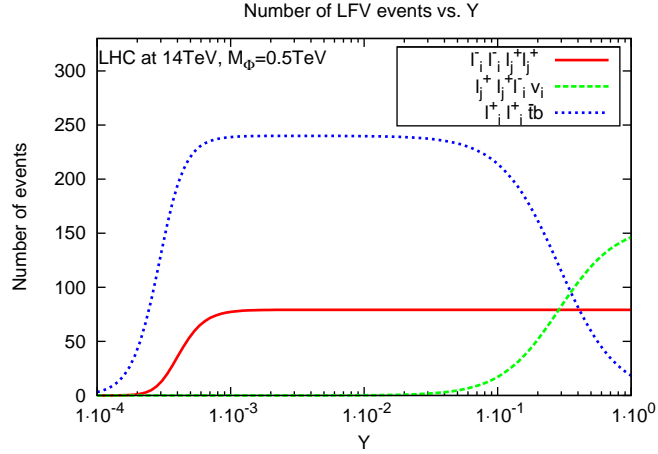}
\end{center}
\caption{Dependence of total number of lepton number and lepton flavor violating final states from decays of scalar pairs at LHC when $\sqrt{S}=14TeV$ and $M_\phi=0.5TeV$ for an integrated luminosity of $100fb^{-1}$ on Yukawa coupling $Y$.}
\label{fig::No14}
\end{figure}

\begin{figure}[tbh]
\begin{center}
\includegraphics[width=9cm]{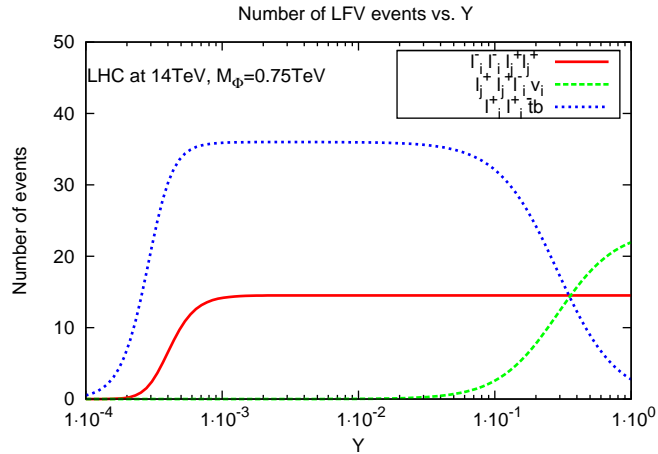}
\end{center}
\caption{Dependence of total number of lepton number and lepton flavor violating final states from decays of scalar pairs at LHC when $\sqrt{S}=14TeV$ and $M_\phi=0.75TeV$ for an integrated luminosity of $100fb^{-1}$ on Yukawa coupling $Y$.}
\label{fig::No14x2}
\end{figure}

Finally, we estimated the dependence of total number of lepton number and lepton flavor violating final states on $Y$ 
for $\sqrt{S}=7TeV$, $\sqrt{S}=14TeV$ when $M_\phi=0.5TeV$ in figures \ref{fig::No7} and \ref{fig::No14} respectively, and for 
$\sqrt{S}=14TeV$ when $M_\phi=0.75TeV$ in figure \ref{fig::No14x2}. In this estimated we assumed an integrated luminosity of 
$100fb^{-1}$ for LHC. It is seen from Fig. \ref{fig::No7} that for $M_\phi=0.5TeV$ and $\sqrt{S}=7TeV$, there can be 
about 20 lepton number violating two lepton plus jet signals observable at LHC for $10^{-3}<Y<10^{-1}$ free from any backgrounds. For $\sqrt{S}=7TeV$, three lepton and four lepton channels have lower event numbers and they are not promising. 
Our final analysis for $M_\phi=0.5TeV$ and $\sqrt{S}=14TeV$ plotted in figure \ref{fig::No14} shows that in this case the number of lepton flavor and number violating final states are more promising. For $10^{-3}<Y<10^{-1}$ there can be 80 four lepton signals violating lepton flavor by four, and also few hundreds of two leptons plus jets final states violating lepton number by two. If $Y\sim 1$, this case is dominated by hundreds of three lepton plus missing energy signals with an observed lepton flavor violation by two. For $\sqrt{S}=14TeV$ and $M_\phi=0.75TeV$, i.e. $f\sim1TeV$ the final number of lepton flavor and number violating events are plotted in figure \ref{fig::No14x2}. For higher values of $M_\phi$, the number of events are reduced almost $60\%$, but still in the reach since they are all free from any SM backgrounds. In this case about $30$ to $40$ signals of two leptons plus jets violating lepton number by two can be expected if $10^{-3}<Y<10^{-1}$. The expected number of events in four lepton or three lepton channels are at the order of $10$, if $10^{-3}<Y<10^{-1}$ or $Y\sim1$ respectively.

\section{Conclusion}

In conclusion, the charged scalars of the littlest Higgs model can be produced via $pp \to \phi^{++}\phi^{-}$,
$pp \to \phi^+\phi^-$ and $pp \to \phi^{++}\phi^{--}$ processes at LHC. Since the production rates are low, the detection of these scalar productions 
can only be done from SM background free final states. For the littlest Higgs model, these signals are the lepton flavor and number violating 
final states arising from the structure of its scalar sector containing  a complex $SU(2)$ group with hypercharge two which can interact with SM particles at tree level. 
In this work we found that depending on the Yukawa coupling $Y$, lepton flavor and number violating final states can be detected at 
LHC. For $Y\sim1$, lepton flavor violation by four in the channel $l_il_i\bar{l}_j\bar{l}_j$ and lepton flavor violation by two 
in the channel $\bar{l}_i\bar{l}_i l_j v_j$ can be observed. If the Yukawa coupling is in the range $10^{-3}<Y<0.1$, lepton number violation by two in the channel 
$\bar{l}_i\bar{l}_i \bar{t} b$ can be observed. In both of these final states the littlest Higgs heavy scalar can be reconstructed from invariant mass distributions of 
same sign and same family lepton pairs. To identify such a signal coming from littlest Higgs model, the mass of the scalar should satisfy $M_\phi\geq 0.5TeV$ due to constraints coming 
from experimental data. If such final states are achieved at LHC, this will be a discriminating discovery for littlest Higgs model among other 
little Higgs models and models containing doubly charged scalars.

%
%
%
%

\clearpage

\begin{thebibliography}{99}
\bibitem{GUT}  G.G. Ross, \textit{``Grand Uniﬁed Theories``} (Addison-Wesley Publishing Company, Reading, MA, (1984); P. Langacker, Phys. Rep. \textbf{72} (1981) 185; 
 H. Georgi, S.L Glashow, Phys. Rev. Lett. 32 (1974) 438; 
A.J. Buras, J. Ellis, M.K. Gaillard, D.V. Nanopoulos, Nucl. Phys. \textbf{B135} (1978) 66.
\bibitem{SUSY}  S. P. Martin, {\tt arXiv:hep-ph/9709356}; M. E. Peskin, {\tt arXiv:0801.1928 [hep-ph]}; 
K. A. Olive, {\tt arXiv:hep-ph/9911307}; 
M. Drees, {\tt arXiv:hep-ph/9611409}; 
H. E. Haber and G. L. Kane, Phys. Rept. \textbf{117} (1985) 75; H. P. Nilles, Phys. Rept. \textbf{110}, 1 (1984);
 A. Signer, J.Phys.G \textbf{G36} (2009) 073002, {\tt arXiv:0905.4630 [hep-ph]}.
\bibitem{ED} N. Arkani-Hamed, S. Dimopoulos, G. Dvali, Phys. Lett. \textbf{B429} (3–4): 263–272(1998), {\tt arXiv:hep-ph/9803315};
N. Arkani-Hamed, S. Dimopoulos, G. Dvali, Phys. Rev. \textbf{D59} 086004(1999), {\tt arXiv:hep-ph/9807344}; I. Antoniadis, N. Arkani-Hamed, S. Dimopoulos, G. Dvali, Phys. Lett. \textbf{B436} (3–4): 257–263(1998), {\tt arXiv:hep-ph/9804398};
M. Shifman, Int.J.Mod.Phys.\textbf{A25} 199-225,2010,{\tt arXiv:0907.3074v2 [hep-ph]}.
\bibitem{331mod} J. Schechter and Y. Ueda, Phys. Rev. \textbf{D8}, 484 (1973); 
J. G. Segr`e and J. Weyers, Phys. Lett. \textbf{B 65}, 243 (1976); 
H. Fritzch and P.Minkowski, Phys. Lett. \textbf{B 63}, 204 (1976); 
B. W. Lee and S. Weinberg, Phys. Rev. Lett. \textbf{38}, 1237 (1977); 
B. W. Lee and R. E. Shrock, Phys. Rev. \textbf{D 17}, 2410 (1978); 
P. Langacker, G.Segr`e and M. Golshani, Phys. Rev. \textbf{D 17}, 1402 (1978); 
M. Singer, J. W. F. Valle and J.Schechter, Phys. Rev. \textbf{D 22}, 738 (1980); 
F. Pisano and V. Pleitez, Phys. Rev. \textbf{D 46} (1992), 410; 
P. H. Frampton, Phys. Rev. Lett. \textbf{69} (1992), 2889; 
R. Foot, H. N. Long and T. A. Tran, Phys. Rev. \textbf{D50}, (1994) R34; 
J. C. Montero, F. Pisano and V. Pleitez, Phys. Rev. \textbf{D47} (1993), 291.
\bibitem{LR}  R. N. Mohapatra and J. C. Pati, Phys. Rev. \textbf{D11} (1975), 566–571; G. Senjanovic and R. N. Mohapatra; Phys. Rev. \textbf{D12} (1975), 1502; A. Adulpravitchai, M. Lindner, A. Merle, and R. N. Mohapatra, Phys. Lett. \textbf{B680} (2009), 476–479 {\tt arXiv/hep-ph:0908.0470}.
\bibitem{BL}L.Basso, S. Moretti, G.M. Pruna, Phys.Rev.\textbf{D83}055014,2011, {\tt arXiv:1011.2612v4 [hep-ph]}.
\bibitem{DoubletM} S. L. Glashow and S. Weinberg, Phys. Rev. \textbf{D 15, 1958 (1977)}; 
V. D. Barger, J. L. Hewett and R. J. N. Phillips, Phys. Rev. \textbf{D 41}, 3421 (1990); 
Y. Grossman, Nucl.Phys. \textbf{B 426}, 355 (1994); 
A. G. Akeroyd, Phys. Lett. \textbf{B 377}, 95 (1996); 
M. Aoki, S. Kanemura, K. Tsumura and K. Yagyu, Phys. Rev. \textbf{D 80}, 015017 (2009); 
V. Barger, H. E. Logan and G. Shaughnessy, Phys. Rev. \textbf{D 79}, 115018 (2009); 
S. Su and B. Thomas; Phys. Rev. \textbf{D 79}, 095014 (2009); 
H. E. Logan and D. MacLennan, Phys. Rev. \textbf{D 79}, 115022 (2009); 
E. Ma, Mod. Phys. Lett. \textbf{A 17}, 535 (2002); 
E. Ma and D. P. Roy, Nucl. Phys. \textbf{B 644}, 290 (2002); 
M. Aoki, S. Kanemura and O. Seto, Phys. Rev. Lett. \textbf{102}, 051805 (2009); Phys. Rev. \textbf{D 80}, 033007(2009); 
M. Aoki, S. Kanemura and K. Yagyu, Phys. Rev. \textbf{D 83}, 075016 (2011).
\bibitem{TripletM}  W. Konetschny and W. Kummer, Phys. Lett. \textbf{B 70}, 433 (1977); R. N. Mohapatra and G. Senjanovic, Phys. Rev. Lett. \textbf{44}, 912 (1980); 
M. Magg and C. Wetterich, Phys. Lett. \textbf{B 94}, 61 (1980); G. Lazarides, Q. Shaﬁ and C. Wetterich, Nucl. Phys. \textbf{B 181}, 287 (1981); 
J. Schechter and J. W. F. Valle, Phys. Rev. \textbf{D 22}, 2227 (1980); T. P. Cheng and L. F. Li, Phys. Rev. \textbf{D 22}, 2860 (1980)
\bibitem{lh1} N. Arkani-Hamed, A.G. Cohen, E. Katz, and A.E. Nelson, JHEP\textbf{0207}(2002)034, {\tt arXiv:hep-ph/0206021}.
\bibitem{lh2} N. Arkani-Hamed \emph{et al.}, JHEP\textbf{0208}(2002)021, {\tt
arXiv:hep-ph/0206020}. 
\bibitem{lhSimple}M. Schmaltz,
Nucl.Phys.Proc.Suppl.\textbf{117}(2003),{\tt arXiv:hep-ph/0210415}.
\bibitem{lhProduct}D.E.Kaplan and M. Schmaltz, JHEP\textbf{0310}(2003)039, {\tt
arXiv:hep-ph/0302049}.
\bibitem{ays1}A. Cagil, M. T. Zeyrek, Phys.Rev. \textbf{D80}055021(2009), {\tt
arXiv:hep-ph/0908.3581}.
\bibitem{ays2}A. Cagil, Nucl.Phys. \textbf{B843} (2011) 46-54 , {\tt arXiv:1010.0102v1 [hep-ph]}, A. Cagil, PoS CHARGED2010 (2010) 033, {\tt arXiv:1102.3898 [hep-ph]}.
\bibitem{ays3}A. Cagil, M.T. Zeyrek, Acta Phys.Polon. \textbf{B42} (2011) 45-60, {\tt arXiv:1010.4139 [hep-ph]}. 
\bibitem{akeyord}A.G. Akeroyd, C-W Chiang, N. Gaur, JHEP 1011 (2010) 005 , {\tt arXiv:1009.2780 [hep-ph]}. 
\bibitem{csaki1}C. Csaki, J. Hubisz, G.D. Kribs, P. Meade, J. Terning, Phys. rev. \textbf{D67} (2003) 115002, {\tt
arXiv:hep-ph/0211124v2}.
\bibitem{perelstein2ew}J. Hubisz, P. Maeda, A. Noble and M. Perelstein,
JHEP\textbf{01}(2006)135.
\bibitem{B1rizzo}J. L. Hewett, F. J. Petriello and T. G. Rizzo, JHEP\textbf{ 0310} (2003) 062 , {\tt
arXiv:hep-ph/0211218}.
\bibitem{Bdawson} M-C. Chen and S. Dawson, Phys.Rev.\textbf{D70}(2004)015003, {\tt arXiv:hep-ph/0311032}.
\bibitem{Bkilian} W. Killian and J. Reuter, Phys.Rev. \textbf{D70} (2004) 015004, {\tt arXiv:hep-ph/0311095}.
\bibitem{Bdias} A.G. Dias, C.A. de S. Pires P.S. Rodrigues da Silva, Phys.Rev.\textbf{D77}(2008)055001, {\tt arXiv:hep-ph/0711.1154}.
\bibitem{Tevatron}T. Aaltonen, et al, CDF Collaboration, Phys.Rev.Lett.99:171802,2007.
\bibitem{B2csaki} C. Csaki, J. Hubisz, G. D. Kribs, P. Maede and J. Terning, Phys.Rev. \textbf{D68} (2003) 035009 , {\tt arXiv:hep-ph/0303236}.
\bibitem{Tparity} H.~C.~Cheng and I.~Low,
  JHEP {\bf 0309}, 051 (2003)
  [arXiv:hep-ph/0308199].
\bibitem{Tparitydarkmatter}H.~C.~Cheng and I.~Low,
  JHEP {\bf 0408}, 061 (2004)
  [arXiv:hep-ph/0405243].
\bibitem{thanlept1}T. Han, H.E. Logan, B. Mukhopadhyaya and R.
Srikanth, Phys.Rev. \textbf{D72} (2005) 053007 , {\tt
arXiv:hep-ph/0505260}.
\bibitem{gaurlept1}S.R. Choudhury, N. Gaur and A. Goyal, Phys.Rev. \textbf{D72} (2005) 097702 , {\tt arXiv:hep-ph/0508146}, S.R. Choudhury \emph{et al.}, Phys.Rev.\textbf{D75}(2007)055011, {\tt arXiv:hep-ph/0612327}.
\bibitem{cinlept_L2yue}C-X. Yue and S. Zhao, Eur.Phys.J.\textbf{C50}(2007)897-903, {\tt arXiv:hep-ph/0701017}.
\bibitem{gaurlept2}F. del Aguila, J. I. Illana, M.D. Jenkins, JHEP 1103:080,2011, {\tt arXiv:1101.2936v2 [hep-ph]}. 
\bibitem{331}D. A. Gutierrez, W. A. Ponce, L.A. Sanchez,  Int.J.Mod.Phys.\textbf{A21} 2217-2235,2006, {\tt arXiv:hep-ph/0511057}.
\bibitem{EXmodelindependent} V. D. Barger, H. Baer, W. Y. Keung and R. J. N. Phillips, Phys. Rev. \textbf{D 26}, 218 (1982); 
J. F. Gunion, J. Grifols, A. Mendez, B. Kayser and F. I. Olness, Phys. Rev. \textbf{D 40}, 1546(1989); 
J. F. Gunion, C. Loomis and K. T. Pitts, eConf C960625, LTH096 (1996) {\tt arXiv:hep-ph/9610237};
 M. Muhlleitner and M. Spira, Phys. Rev. \textbf{D 68}, 117701 (2003); 
T. Han, B. Mukhopadhyaya, Z. Si and K. Wang, Phys. Rev. \textbf{D 76}, 075013 (2007); 
K. Huitu, J. Maalampi, A. Pietila and M. Raidal, Nucl. Phys. \textbf{B 487}, 27 (1997); 
J. Maalampi and N. Romanenko, Phys. Lett. \textbf{B 532}, 202 (2002); 
B. Dion, T. Gregoire, D. London, L. Marleau and H. Nadeau, Phys. Rev. \textbf{D 59}, 075006(1999); 
A. G. Akeroyd and M. Aoki, Phys. Rev. \textbf{D 72}, 035011 (2005).
\bibitem{mphi}D. E. Acosta et al. [CDF Collaboration], Phys. Rev. Lett. \textbf{93}, 221802 (2004); V. M. Abazov et al. [D0 Collaboration], Phys. Rev. Lett. \textbf{93}, 141801 (2004); 
V. M. Abazov et al. [D0 Collaboration], Phys. Rev. Lett. \textbf{101}, 071803 (2008); 
T. Aaltonen et al. [The CDF Collaboration], Phys. Rev. Lett. \textbf{101}, 121801 (2008).
\bibitem{ref1Charged} C-X. Yue, S. Zhao and W. Ma, Nucl.Phys.\textbf{B784}, 36-48,2007, {\tt arXiv:0706.0232v3 [hep-ph]}.
\bibitem{thanrev}T. Han, H. E. Logan, B. McElrath and L-T. Wang, Phys.Rev. \textbf{D67}(2003)095004, {\tt arXiv:hep-ph/0301040}.
\bibitem{neutrinomass}G.L. Fogli \emph{et al}, Phys. Rev. D\textbf{70},
113003(2004); M. Tegmark \emph{et al}, SDSS Collabration, Phys. Rev.
D\textbf{69}, 103501(2004).
\bibitem{aysbook} A. Cagil, \textit{``New Higgs Scalars of the Littlest Higgs Model``}, LAP Lambert Academic Publishing, ISBN:9783846554395, 2011.
\bibitem{pdg}K. Nakamura et al. (Particle Data Group), Journal of Physics G37, 075021 (2010). 
\bibitem{LHClum} ATLAS collaboration, Eur.Phys.J.C71:1630,2011, {\tt arXiv:1101.2185v1 [hep-ex]}.
\end{thebibliography}
\end{document}